\documentclass[aps,prl,twocolumn,showpacs,preprintnumbers,superscriptaddress,groupedaddress]{revtex4}  
\usepackage{graphicx}  
\usepackage{dcolumn}   
\usepackage{bm}        
\usepackage{slashed}
\usepackage{amssymb} \usepackage{amsmath}   
\usepackage{xcolor}

\def\sq[#1,#2]{\left[#1\,#2\right]}
\def\an[#1,#2]{\left\langle#1\,#2\right\rangle}
\def\spab[#1,#2,#3]{\left\langle#1|#2|#3\right]}
\newcommand{\vev}[1]{\left\langle #1 \right\rangle}
\newcommand{\ord}[1]{{\scriptscriptstyle (#1)}}

\begin{document}
\preprint{}
\title{The Schwarzschild Black Hole from Perturbation Theory to all Orders}
\author{Poul H. Damgaard}\email{phdamg@nbi.dk}
\affiliation{Niels Bohr  International  Academy,
The Niels Bohr Institute, Blegdamsvej 17,
DK-2100 Copenhagen \O, DENMARK}
\author{Kanghoon Lee}\email{kanghoon.lee1@gmail.com}
\affiliation{Asia Pacific Center for Theoretical Physics, Postech,
Pohang 37673, Korea}
\affiliation{Department of Physics, Postech, Pohang 37673, Korea}
\date{\today}

\begin{abstract}
Applying the quantum field theoretic perturbiner approach to Einstein gravity, we compute the metric of a Schwarzschild black hole order by order in perturbation theory. Using recursion, this calculation can be carried out in de Donder gauge to all orders in Newton's constant. The result is a geometric series which is convergent outside a disk of finite radius, and it agrees within its region of convergence with the known de Donder gauge metric of a Schwarzschild black hole. It thus provides a first all-order perturbative computation in Einstein gravity with a matter source, and this series converges to the known non-perturbative expression in the expected range of convergence.
\end{abstract}
\maketitle

\section{Introduction}

The recent surge of interest in gravitational-wave physics associated with the merging of two black holes has led to a number of powerful new analytical approaches to Einstein gravity.
This has put the post-Minkowskian expansion, the perturbative evaluation of physical observables as a power series in Newton's constant $G$ around flat Minkowski space, into focus.
Because of its close connection to the classical limit of quantum field theoretical scattering amplitudes in gravity \cite{Bjerrum-Bohr:2018xdl,Cheung:2018wkq,Kosower:2018adc,Cristofoli:2019neg}, rapid progress has led to the exact analytical evaluation of physical observables up to (and including) fourth order in the perturbative expansion in the coupling constant $G$ \cite{Bern:2019nnu,Bern:2019crd,Kalin:2019rwq,Bjerrum-Bohr:2019kec,Damour:2019lcq,DiVecchia:2020ymx,Damour:2020tta,DiVecchia:2021ndb,DiVecchia:2021bdo,Herrmann:2021tct,Bjerrum-Bohr:2021vuf,Bjerrum-Bohr:2021din,Damgaard:2021ipf,Brandhuber:2021eyq,Brandhuber:2021bsf,Kalin:2020mvi,Kalin:2020fhe,Mogull:2020sak,Jakobsen:2021smu,Mougiakakos:2021ckm,Riva:2021vnj,Bern:2021dqo,Dlapa:2021npj,Bern:2021yeh,Adamo:2022ooq,Dlapa:2023hsl,Jakobsen:2023ndj,Damgaard:2023vnx,Damgaard:2023ttc,Jakobsen:2023hig,DiVecchia:2023frv,Driesse:2024xad}, also in some cases including low orders in the spin multipole expansion. Such a successful program, which has already reached fifth order in the expansion \cite{Driesse:2024xad}, prompts the question as to whether the perturbative series expansion in Einstein gravity is convergent. One of the first examinations of this issue dates back to ref. \cite{Christodoulou}; see also ref. \cite{Damour:1990rm}. Here we shall investigate the question of convergence of the post-Minkowskian expansion in its simplest possible setting: that of vacuum solutions to the Einstein equations in the presence of a point-like source. The surprising difficulty in deriving the metric perturbatively in this context was noted first by Florides and Synge \cite{Synge}. Low-order perturbative calculations of the gravitational metric of such a non-spinning black hole have since been pursued by means of several different quantum field theoretical formalisms \cite{Duff:1973zz,Bjerrum-Bohr:2002fji,Goldberger:2004jt,Neill:2013wsa,KoemansCollado:2018hss,Jakobsen:2020ksu,Mougiakakos:2020laz,DOnofrio:2022cvn}. Eventually, and typically at quite low orders, they run into a problem posed by the complexity of the perturbatively expanded Einstein-Hilbert action. This pushes the calculations to prohibitively complicated levels beyond the first few orders. In the amplitude-based approach to general relativity this problem is partly circumvented by the use of on-shell unitarity methods. Such techniques are, however, not immediately amenable to the off-shell situation needed for the calculation of the gravitational metric itself. Instead, as we shall demonstrate below, a judicious choice of variables, combined with off-shell recursion relations of the current as well as relevant loop integrals, can lead to drastic simplifications.

Our tool shall be the perturbiner approach to quantum field theory \cite{Rosly:1996vr,Rosly:1997ap,Mafra:2016ltu,Mizera:2018jbh,Gomez:2021shh,Gomez:2022dzk,Cho:2021nim,Lee:2022aiu,Cho:2022faq,Adamo:2023cfp,Lee:2023zuu,Cho:2023kux,Tao:2023yxy}. This method provides an alternative path towards scattering amplitudes and in the case of the gravitational field yields an analog of Berends-Giele \cite{Berends:1987me} off-shell recursion relations for gravity. The perturbiner method combines very well with the post-Minkowskian expansion of gravity \cite{Cho:2021nim,Lee:2022aiu,Cho:2022faq,Adamo:2023cfp,Lee:2023zuu,Cho:2023kux}, leading to the pertinent two-body scattering amplitudes in a manner that is closely linked to the classical equations of motion or, at loop level, the Schwinger-Dyson equations. Combining the perturbiner method with good variables for the gravitational field is the first key ingredient towards obtaining an all-order result in perturbation theory. This still leaves us with the problem that the Einstein-Hilbert action expanded in coupling $G$ becomes immensely complicated beyond lowest orders. An important observation is that the complexity arises to a large degree from iterations of lower-point vertices and this can be used to advantage, leading us to the second key ingredient of our approach which is to expand the Einstein-Hilbert action perturbatively in a doubled set of fields. The required constraint among the two sets of fields is imposed afterward. Finally, we will need to evaluate a series of momentum-space integrals. Our third key observation is that these integrals reduce to generalized 1-loop bubble diagrams. They are then all solvable by iteration. The various new recursive identities that have been discovered in the theory of scattering amplitudes rely on the observation that each new order can be built up from combinations of lower-order terms. This is also the mechanism behind the present calculation and it illustrates how modern quantum field theory methods can provide a surprising new avenue to general relativity. 

\section{The metric}

The Schwarzschild metric in harmonic gauge is given by
\begin{equation}
\begin{aligned}
    \mathrm{d} s^{2} &= - N \mathrm{d} t^{2} + \Bigg[F \delta^{ij} 
  + \frac{G^{2}M^{2}}{N} \frac{x^{i}x^{j}}{r^{4}}\Bigg] \mathrm{d}x^{i} \mathrm{d}x^{j}\,,
\end{aligned}\label{}
\end{equation}
where $r^{2} = x_{1}^{2}+x_{2}^{2} +x_{3}^{2}$, and 
\begin{equation}
  N(r) = \frac{r-GM}{r + GM}\,, 
  \qquad 
  F(r) = \frac{(r+GM)^{2}}{r^{2}}\,.
\label{}\end{equation}
We choose to work with Landau-Lifshitz variables and expand perturbatively around flat Minkowski 
space as follows:
\begin{equation}
  \mathfrak{g}^{\mu\nu} \equiv \sqrt{-g} g^{\mu\nu} = \eta^{\mu\nu} - h^{\mu\nu} \,.
\label{metric_perturbation}\end{equation}
%
%
%
With this choice of variables, the fluctuation field $h^{\mu\nu}$ reads
\begin{equation}
h^{00} = -1 + \frac{\left(r+G M\right)^3}{r^2 \left(r-G M\right)} ~~,~~~ 
 h^{ij} = \frac{G^2 M^2 x^{i}x^{j}}{r^4}\,.
\label{metric_components}
\end{equation}
Expanding $h^{00}$ as a Laurent series around 1 we get
\begin{eqnarray}
h^{00} &=& 8\left(1 - \frac{GM}{r}\right)^{-1} - 8 - \frac{4GM}{r} - \frac{G^2M^2}{r^2} \cr
&=& \frac{4GM}{r} + \frac{7G^2M^2}{r^2}  + \frac{8G^3M^3}{r^3} \ldots
\label{Laurent_series}
\end{eqnarray}
Except for the few low-order corrections $h^{00}$ is thus a geometric series in $GM/r$ with fixed 
coefficient 8, while $h^{ij}$ truncates at second order.
The fluctuation field $h^{\mu\nu}$ satisfies the de Donder gauge 
condition $\partial_{\mu} (\sqrt{-g} g^{\mu\nu}) = \partial_{\mu} h^{\mu\nu} = 0$  
which is equivalent to the harmonic gauge condition in these variables. 
%
%

\section{The Perturbiner method}

We now set up the post-Minkowskian perturbative expansion for the Schwarzschild black hole metric
as defined in eq. (\ref{metric_perturbation}).
We do this by generalizing the usual perturbiner method for scattering amplitudes to solutions of the perturbative Einstein equations. Specifically, we wish to generate the perturbative vacuum solution of Einstein gravity for a point-like source at the origin, corresponding to a Schwarzschild black hole. 

The action is thus given by
\begin{equation}
    S = \int \mathrm{d}^{4}x \bigg[ \frac{1}{2\kappa^{2}} \sqrt{-g} R + \frac{1}{2} j_{\mu\nu}(x) g^{\mu\nu}(x) \bigg] \,, 
\label{EHaction}\end{equation}
where $R$ is the Ricci scalar and the source $j_{\mu\nu}(x)$ is  
\begin{equation}
  j_{\mu\nu}(x) = M v_{\mu} v_{\nu} \delta^{3}(\boldsymbol{x})\,, \qquad v_{\mu} = (-1,0,0,0)\,.
\label{delta_function_source}\end{equation}
Here $\boldsymbol{x}$ denotes the position 3-vector. 
%
%
Let us consider the one-point function of $h^{\mu\nu}_{x}\equiv h^{\mu\nu}(x)$ in the presence of the source 
$j_{\mu\nu}$,
\begin{equation}
  \mathsf{h}^{\mu \nu}_{x} = \vev{0|h^{\mu\nu}_{x}|0}_{j} =\frac{\delta W[j]}{\delta j_{x}^{\mu \nu}}\,,
\label{}\end{equation}
where $W[j_{\mu\nu}]$ is the generating functional
\begin{equation}
  e^{W[j_{\mu\nu}]} \equiv \int \mathcal{D} h^{\mu \nu} \exp \left[\frac{i}{\hbar}S[h^{\mu\nu},j_{\mu\nu}]\right]\,.
\label{}\end{equation}
Note that $\mathsf{h}^{\mu\nu}_{x}$ satisfies the classical equations of motion at tree level. 
Since we are considering a time-independent solution, we restrict $\mathsf{h}^{\mu\nu}(x) = \mathsf{h}^{\mu\nu}(\boldsymbol{x})$.

Expanding the generating functional in powers of $j_{\mu\nu}$, $\mathsf{h}^{\mu\nu}_{x}$ is given
in terms of connected correlation functions,
\begin{equation}
\begin{aligned}
    \mathsf{h}^{\mu\nu}_{\boldsymbol{x}} &=\sum_{n=1}^{\infty} \frac{1}{n !} \int_{y_{1}, \cdots, y_{n}} 
\!\!\!\! \big\langle 0\big|T\big[h_{x}^{\mu\nu} h_{y_{1}}^{\kappa_{1}\lambda_{1}} \cdots h_{y_{n}}^{\kappa_{n}\lambda_{n}}\big]\big| 0\big\rangle_{c} 
 \\
 &\quad
 \times\frac{i j_{y_{1}}^{\kappa_{1}\lambda_{1}}}{\hbar} \cdots \frac{i j_{y_{n}}^{\kappa_{n}\lambda_{n}}}{\hbar}\,,
\end{aligned}\label{}
\end{equation}
%
%
%
where
\begin{equation}
  \int_{x, y \cdots}=\int \mathrm{d}^{d} x \mathrm{d}^{d} y \cdots \text { and } \int_{p, q, \cdots}=\int \frac{\mathrm{d}^{d} p}{(2 \pi)^{d}} \frac{\mathrm{d}^{d} q}{(2 \pi)^{d}} \cdots\,.
\label{}\end{equation}
%
%
%
Hereafter, we will ignore the position of indices while keeping the summation convention for repeated indices.

Next, we substitute $j_{\mu\nu}$ as defined in \eqref{delta_function_source}. It is useful to represent the delta function as
\begin{equation}
  j^{\mu\nu}_{\boldsymbol{x}} = Mv^{\mu} v^{\nu} \int_{\boldsymbol{\ell}} e^{-i \boldsymbol{\ell}\cdot \boldsymbol{x}}\,.
\label{}\end{equation}
Then $\mathsf{h}^{\mu\nu}_{x}$ reduces to
\begin{equation}
\begin{aligned}
  \mathsf{h}^{\mu\nu} (\boldsymbol{x}) &= 
  \sum_{n=1}^{\infty} \frac{1}{n !}\int_{\boldsymbol{\ell}_{1},\boldsymbol{\ell}_{2},\cdots,\boldsymbol{\ell}_{n}} 
  J^{\mu\nu}_{\boldsymbol{\ell}_{1}\boldsymbol{\ell}_{2}\cdots\boldsymbol{\ell}_{n}} e^{-i\boldsymbol{\ell}_{12\cdots n}\cdot \boldsymbol{x}}\,,
\end{aligned}\label{}
\end{equation}
where 
%
\begin{equation}
  J^{\mu\nu}_{\boldsymbol{\ell}_{1}\boldsymbol{\ell}_{2}\cdots\boldsymbol{\ell}_{n}} = \left(\frac{iM}{\hbar}\right)^{n}\big\langle 0\big|T\big[\tilde{h}_{-\boldsymbol{\ell}_{12\cdots n}}^{\mu\nu} \tilde{h}_{\boldsymbol{\ell}_{1}}^{00} \cdots \tilde{h}_{\boldsymbol{\ell}_{n}}^{00}\big]\big| 0\big\rangle_{c}\,,
\label{}\end{equation}
and $\boldsymbol{\ell}_{12\cdots n} = \boldsymbol{\ell}_{1}+\boldsymbol{\ell}_{2}+\cdots +\boldsymbol{\ell}_{n}$. Here $\tilde{h}_{\ell}^{\mu\nu}$ is the Fourier transform of $h_x^{\mu\nu}$.
It is convenient to shift the loop momentum $\boldsymbol{\ell}_{1} \to - \boldsymbol{\ell}_{12\cdots n}$ so that
\begin{equation}
  \mathsf{h}^{\mu\nu}(\boldsymbol{x}) = 
  \sum_{n=1}^{\infty} \int_{\boldsymbol{\ell}_{1}} e^{i\boldsymbol{\ell}_{1}\cdot \boldsymbol{x}} J^{\mu\nu}_{\ord{n}| \boldsymbol{\ell}_{1}} \,,
\end{equation}
where
\begin{equation}
  J^{\mu\nu}_{\ord{n}|\boldsymbol{\ell}_{1}} = \int_{\boldsymbol{\ell}_{2},\cdots,\boldsymbol{\ell}_{n}}  \frac{1}{(n-1)!} J^{\mu\nu}_{-\boldsymbol{\ell}_{12\cdots n}\boldsymbol{\ell}_{2}\cdots\boldsymbol{\ell}_{n}}\,.
\label{}\end{equation}

We denote the number of loop momenta of an off-shell current by its `rank'. For instance, the rank of $J^{\mu\nu}_{\ell_{1}\cdots\ell_{n}}$ is $n$. We then expand 
$\mathsf{h}^{\mu\nu}$ according to this rank, $i.e.$
\begin{equation}
  \mathsf{h}^{\mu\nu} = \sum_{n=0}^{\infty} G^{n}\mathsf{h}^{\mu\nu}_{\ord{n}} ~~,~~~ \mathsf{h}^{\mu\nu}_{\ord{n}} \equiv \int_{\boldsymbol{\ell}} J^{\mu\nu}_{\ord{n}|\boldsymbol{{\ell}}} e^{i \boldsymbol{\ell}\cdot \boldsymbol{x}}\,.
\label{expansion_h}\end{equation}
%
%
%

%
%


We now proceed to organize the perturbative expansion of the Einstein-Hilbert action. Because $\mathsf{h}^{\mu\nu}$ satisfies the classical equations of motion at tree level, we substitute \eqref{metric_perturbation}
into \eqref{EHaction} and expand the equation of motion in powers of $G$,
%
\begin{equation}
\Box \mathsf{h}^{\mu\nu} = -2 j^{\mu\nu} - \sum_{n=1}^{\infty} \tau^{\mu\nu}_{[n]} ~,
\end{equation}
where the $\tau^{\mu\nu}_{[n]}$'s arise in the well-known manner from gravitational self-interactions (and they will be given in
very compact forms below). This expansion is by construction also ordering the number of fields $\mathsf{h}^{\mu\nu}$ so that $\tau^{\mu\nu}_{[n]}$  contains $(n+1)$ fields.
The off-shell recursion relations are now obtained by substituting the perturbiner expansion. 
The initial condition of the recursion is given by the rank-1 current, 
%
\begin{equation}
  \Delta \mathsf{h}^{\mu\nu}_{\ord{1}} = 
-2 j^{\mu\nu}
= -2M v^{\mu} v^{\nu} \int_{\boldsymbol{k}} e^{i \boldsymbol{k}\cdot \boldsymbol{x}}\,,
\end{equation}
where $\Delta$ is the Laplacian. 
%
%
Substituting the perturbiner expansion \eqref{expansion_h} for $\mathsf{h}^{\mu\nu}_{\ord{1}}$
this provides us with the initial condition for the off-shell recursion relation
\begin{equation}
  J^{\mu\nu}_{\ord{1}|\boldsymbol{\ell}} 
  = \frac{16 \pi M}{|\boldsymbol{\ell}|^{2}} v^{\mu} v^{\nu}\,,
\label{rank_one}\end{equation}
or 
\begin{equation}
  J^{00}_{\ord{1}|\boldsymbol{\ell}} = \frac{16 \pi M}{|\boldsymbol{\ell}|^{2}}\,,
  \qquad
  J^{0i}_{\ord{1}|\boldsymbol{\ell}} = 0\,,
  \qquad
  J^{ij}_{\ord{1}|\boldsymbol{\ell}} =0\,.
\label{initial_condition}\end{equation}
%
We now proceed to rank-2 level. According to the initial condition in \eqref{initial_condition}, the only nonvanishing component of $\mathsf{h}_{\ord{1}}^{\mu\nu}$ is $\mathsf{h}_{\ord{1}}^{00}$. 
%
%
%
%
Inserting the equations of motion to this order, we get the off-shell recursion relations at rank-2 from the
perturbiner expansion,
%
\begin{equation}
\begin{aligned}
  J^{00}_{\ord{2}|-\boldsymbol{\ell}_{1}}&= \frac{1}{|\boldsymbol{\ell}_{1}|^{2}} \int_{\boldsymbol{\ell_{2}}} \bigg[
    \frac{5}{4} |\boldsymbol{\ell}_{2}|^{2} - \frac{7}{8} \boldsymbol{\ell}_{12} \cdot \boldsymbol{\ell}_{2} \bigg] J^{00}_{\ord{1}|-\boldsymbol{\ell}_{12}} J^{00}_{\ord{1}|\boldsymbol{\ell}_{2}}\,,
  \\
  J^{ij}_{\ord{2}|-\boldsymbol{\ell}_{1}} &= \frac{1}{|\boldsymbol{\ell}_{1}|^{2}} \int_{\boldsymbol{\ell_{2}}}\bigg[
      \frac{\ell^{(i}_{12} \ell^{j)}_{2}}{4} 
    - \frac{\delta^{ij}\boldsymbol{\ell}_{12} \cdot \boldsymbol{\ell}_{2}}{8}
  \bigg] 
  J^{00}_{\ord{1}|-\boldsymbol{\ell}_{12}} J^{00}_{\ord{1}|\boldsymbol{\ell}_{2}} \,,
\end{aligned}\label{}
\end{equation}
which shows how $J^{ij}$-components build up from the recursion. Performing the integrals, we get
\begin{equation}
\begin{aligned}
  &J^{00}_{\ord{2}|\boldsymbol{\ell}} = \frac{14\pi^{2} M^{2}}{|\boldsymbol{\ell}|} ,~~
  J^{ij}_{\ord{2}|\boldsymbol{\ell}} = \pi^{2} M^{2}\left[
    \frac{\delta^{ij}}{|\boldsymbol{\ell}|}
  - \frac{\ell^{i}\ell^{j}}{|\boldsymbol{\ell}|^{3}}
  \right]\,.
\end{aligned}\label{}
\end{equation}
%
Computational tedium would seem to prevent us from proceeding to rank-3 level (and beyond). However,
we do need the explicit evaluation for rank-3 in order to establish an all-order result by induction. This
is the first level at which we need to regularize the integrals and we choose to work with dimensional
regularization by going to $d = 3 -2\epsilon$ dimensions at intermediate steps.
We first find $\mathsf{h}^{\mu\nu}_{\ord{3}}$ from the explicit forms of $\tau^{\mu\nu}_{[1]}$ and 
$\tau^{\mu\nu}_{[2]}$ by using the expansion of $\mathsf{h}^{\mu\nu}$ in \eqref{expansion_h}. This gives
\begin{eqnarray}
 J^{00}_{\ord{3}|\boldsymbol{\ell}} &=& 
  \frac{M^3 \pi ^{\epsilon +\frac{3}{2}}}{2^{-4 \epsilon -1}|\boldsymbol{\ell}|^{2 \epsilon }} \frac{(\epsilon  (2 \epsilon +5)+6) 
  \Gamma [1-2 \epsilon] }{ \Gamma \left[\frac{5}{2}-2 \epsilon \right]}\Gamma [\epsilon ] 
  \cr
  J^{ij}_{\ord{3}|\boldsymbol{\ell}} &=& 
  \frac{M^3 \pi ^{\epsilon +\frac{3}{2}}}{4^{-2 \epsilon -1} |\boldsymbol{\ell}|^{2 \epsilon}} \frac{\delta^{ij} \epsilon \Gamma [3-2 \epsilon ] \Gamma[\epsilon -1] }{(2 \epsilon -1) \Gamma[\frac{5}{2}-2 \epsilon ]}\,.
\end{eqnarray}
Fourier-transforming all of these results according to \eqref{expansion_h} and going to $d=3$ dimensions, we obtain
\begin{equation}
  \mathsf{h}^{\mu\nu}_{\ord{1}} = \frac{4M}{r} v^{\mu}v^{\nu}\,,
\label{}\end{equation}
so that
\begin{equation}
  \mathsf{h}^{00}_{\ord{1}} = \frac{4M}{r}\,, 
  \qquad 
  \mathsf{h}^{0i}_{\ord{1}} = \mathsf{h}^{ij}_{\ord{1}} = 0\,.
\label{solution_tree}\end{equation}
at rank-1 level. At rank-2,
\begin{equation}
\begin{aligned}
  \mathsf{h}^{00}_{\ord{2}} &= 
  \int_{\boldsymbol{\ell}} e^{i \boldsymbol{\ell}\cdot \boldsymbol{x}} 
  	J^{00}_{\ord{2}|\boldsymbol{\ell}} = \frac{7M^{2}}{r^{2}}\,, 
  \\
  \mathsf{h}^{ij}_{\ord{2}} &= 
  \int_{\boldsymbol{\ell}} e^{i \boldsymbol{\ell} \cdot \boldsymbol{x}} 
  	J^{ij}_{\ord{2}|\boldsymbol{\ell}} = \frac{M^{2}x^{i}x^{j}}{r^{4}}\,.
\end{aligned}\label{}
\end{equation}
while at rank-3 we get
\begin{equation}
\begin{aligned}
  \mathsf{h}^{00}_{\ord{3}} = \frac{8M^{3}}{r^{3}}\,,
  \qquad 
  \mathsf{h}^{ij}_{\ord{3}} = 0\,.
\end{aligned}\label{}
\end{equation}
These results match the metric of \eqref{metric_perturbation} to ${\mathcal O}(G^3)$.

\section{Iterated loop integrals}

As an intermediate step towards extending the above analysis to all orders we now demonstrate that
the loop integrals to any order are iterated one-loop bubble integrals, and hence solvable recursively.
We use the fact that a generic one-loop bubble integral is of the form
%
\begin{equation}
  \int_{\boldsymbol{\ell}} \frac{\ell^{\rho_{1}} \cdots \ell^{\rho_{m}}}{|\boldsymbol{\ell}|^{\alpha_{1}}|\boldsymbol{\ell}+\boldsymbol{p}|^{\alpha_{2}}} 
  = 
  \frac{M^{\rho_{1}\cdots \rho_{m}}[\boldsymbol{p}]}{|\boldsymbol{p}|^{\beta}} \,,
\label{bubble_integrals}\end{equation}
%
and apply it recursively to integrals of the perturbiner method,
\begin{equation}
\begin{aligned}
  \mathcal{J}^{\mu\nu}_{\ord{n}|\boldsymbol{\ell}_{1}} =& \int_{\boldsymbol{\ell}_{2},\cdots, \boldsymbol{\ell}_{i}} 
  I^{\mu\nu,\kappa_{1}\lambda_{1},\kappa_{2}\lambda_{2}, \cdots, \kappa_{i}\lambda_{i}}[\boldsymbol{\ell}_{1},\cdots, \boldsymbol{\ell}_{i}] \cr
  \quad\times &
  J^{\kappa_{1}\lambda_{1}}_{\ord{N_{1}}|-\boldsymbol{\ell}_{12\cdots n}} 
  J^{\kappa_{2}\lambda_{2}}_{\ord{N_{2}}|\boldsymbol{\ell}_{2}} 
  \cdots 
  J^{\kappa_{i}\lambda_{i}}_{\ord{N_{i}}|\boldsymbol{\ell}_{i}}\,,
\end{aligned}\label{structure_recursion}
\end{equation}
where $n= N_{1}+N_{2}+ \cdots N_{i}$. For off-shell currents of a single propagator,
\begin{equation}
  J^{\mu\nu}_{\ord{m}|\boldsymbol\ell} = \frac{N^{\mu\nu}_{\ord{m}}[\boldsymbol\ell]}{|\boldsymbol\ell|^{\alpha_{m}}}\,,
\label{assumption_currents}\end{equation}
we will now show that $\mathcal{J}^{\mu\nu}_{\ord{n}|\boldsymbol{\ell}_{1}}$ in \eqref{structure_recursion} again contains a single propagator by using \eqref{assumption_currents} and \eqref{bubble_integrals}.
For currents \eqref{assumption_currents}, we thus have
\begin{equation}
\begin{aligned}
  \mathcal{J}^{\mu\nu}_{\ord{n}|\boldsymbol{\ell}_{1}} &= \int_{\boldsymbol{\ell}_{2},\cdots, \boldsymbol{\ell}_{n}} \frac{N^{\mu\nu}[\boldsymbol{\ell}_{1},\cdots,\boldsymbol{\ell}_{n}]}{|\boldsymbol{\ell}_{12\cdots n}|^{\alpha_{1}} |\boldsymbol{\ell}_{2}|^{\alpha_{2}}\cdots |\boldsymbol{\ell}_{n}|^{\alpha_{n}}}\,,
\end{aligned}\label{general_loop_integral1}
\end{equation}
where the numerator is polynomial in the arguments.
We now perform the $\boldsymbol{\ell}_{2}$-integration to get
%
%
%
\begin{equation}
\begin{aligned}
  \mathcal{J}^{\mu\nu}_{\ord{n}|\boldsymbol{\ell}_{1}} 
  &=
  \int_{\boldsymbol{\ell}_{3},\cdots, \boldsymbol{\ell}_{n}} 
  \frac{N_{2}^{\mu\nu}[\boldsymbol{\ell}_{1},\boldsymbol{\ell}_{3}\cdots,\boldsymbol{\ell}_{n}]}{|\boldsymbol{\ell}_{13\cdots n}|^{\beta_{2}} |\boldsymbol{\ell}_{3}|^{\alpha_{3}}\cdots |\boldsymbol{\ell}_{n}|^{\alpha_{n}}} \,,
\end{aligned}\label{}
\end{equation}
with a new numerator $N_2^{\mu\nu}$.
%
%
Since the structure of the integrand is the same as before integration, we may repeat the previous procedure and perform the $\boldsymbol{\ell}_{3}$ integral to get
\begin{equation}
  \mathcal{J}^{\mu\nu}_{\ord{n}|\boldsymbol{\ell}_{1}} =  
  \int_{\boldsymbol{\ell}_{4},\cdots, \boldsymbol{\ell}_{n}} \frac{N_{3}^{\mu\nu}[\boldsymbol{\ell}_{1},\boldsymbol{\ell}_{4},\cdots,\boldsymbol{\ell}_{n}]}{|\boldsymbol{\ell}_{14\cdots n}|^{\beta_{3}}|\boldsymbol{\ell}_{4}|^{\alpha_{4}} \cdots |\boldsymbol{\ell}_{n}|^{\alpha_{n}}}\,,
\label{}\end{equation}
with numerator $N_3^{\mu\nu}$.
This continues until all loop integrals are exhausted, 
leading to
\begin{equation}
\begin{aligned}
  \mathcal{J}^{\mu\nu}_{\ord{n}|\boldsymbol{\ell}_{1}}
  &= \frac{N^{\mu\nu}_{n}[\boldsymbol{\ell}_{1}]}{|\boldsymbol{\ell}_{1}|^{\beta_{n}}}\,,
\end{aligned}\label{}
\end{equation}
where $N_n^{\mu\nu}$ follows recursively in the manner indicated.

\section{Perturbative solution to all orders}

So far we have exploited off-shell recursion relations and iterated momentum integrals.
Our third step towards deriving an all-order result is to encode the fact that the metric and its inverse
expanded around flat Minkowski space are related by a geometric series which again is of iterative form.
We implement this by first considering $\mathfrak{g}^{\mu\nu}$ and its inverse $\mathfrak{g}_{\mu\nu}$ as
independent fields, introducing an auxiliary field $\tilde{\mathfrak{g}}_{\mu\nu}$ whose value is fixed to be
the inverse of $\mathfrak{g}^{\mu\nu}$: $\mathfrak{g}^{\mu\rho}\tilde{\mathfrak{g}}_{\rho\nu} = \delta^{\mu}{}_{\nu}$.
We thus introduce perturbations of both $\mathfrak{g}^{\mu\nu}$ and $\tilde{\mathfrak{g}}_{\mu\nu}$ \cite{Gomez:2021shh},
\begin{equation}
\begin{aligned}
  \mathfrak{g}^{\mu\nu} = \eta^{\mu\nu} - \mathsf{h}^{\mu\nu}\,,
  \qquad
  \tilde{\mathfrak{g}}_{\mu\nu} = \eta_{\mu\nu} + \tilde{\mathsf{h}}_{\mu\nu}\,,
\end{aligned}\label{}
\end{equation}
which we expand in Newton's constant $G$, 
\begin{equation}
  \mathsf{h}^{\mu\nu} = \sum_{n=1}^{\infty} G^{n} \mathsf{h}^{\mu\nu}_{\ord{n}} \,, 
  \qquad
  \tilde{\mathsf{h}}^{\mu\nu} = \sum^{\infty}_{n=1} G^{n} \tilde{\mathsf{h}}^{\mu\nu}_{\ord{n}}\,,
\label{}\end{equation}
%
%
%
with the constraint
\begin{equation}
  \tilde{\mathsf{h}}^{\mu\nu}_{\ord{n}} = 
  \mathsf{h}^{\mu\nu}_{\ord{n}} + \sum_{m=1}^{n-1}\tilde{\mathsf{h}}^{\mu\rho}_{\ord{n-m}} \mathsf{h}^{\rho\nu}_{\ord{m}}\,.
\label{}\end{equation}
We also introduce off-shell currents for $\tilde{\mathsf{h}}^{\mu\nu}_{\ord{n}}$ 
\begin{equation}
  \tilde{\mathsf{h}}^{\mu\nu}_{\ord{n}} = \int_{\boldsymbol{\ell}} e^{i \boldsymbol{\ell}\cdot \boldsymbol{x}} \tilde{J}^{\mu\nu}_{\ord{n}|\boldsymbol{\ell}}\,,
\label{}\end{equation}
where $\tilde{J}^{\mu\nu}_{\ord{n}}$ satifies the following recursion relation:
\begin{equation}
\begin{aligned}
  \tilde{J}^{\mu\nu}_{\ord{n}|\boldsymbol{p}} = 
    J^{\mu\nu}_{\ord{n}|\boldsymbol{p}} 
  + \sum_{m=1}^{n-1} \int_{\boldsymbol{\ell}} \tilde{J}^{\mu\rho}_{\ord{n-m}|\boldsymbol{p}-\boldsymbol{\ell}} J^{\rho\nu}_{\ord{m}|\boldsymbol{\ell}}\,.
\end{aligned}\label{}
\end{equation}
%
%
%
We will now show that the expansion of $\mathsf{h}^{\mu\nu}$ is 
\begin{equation}
\begin{aligned}
  \mathsf{h}^{00}_{\ord{n}} &= \frac{8M^{n}}{r^{n}}\,,
  &\qquad
  \mathsf{h}^{ij}_{\ord{n}} &= 0 \,, 
  \qquad 
  \text{for}~ n\geq3\,,
\end{aligned}\label{}
\end{equation}
while $\tilde{\mathsf{h}}^{00}$ and $\tilde{\mathsf{h}}^{ij}$ are given by
\begin{equation}
\begin{aligned}
  \tilde{\mathsf{h}}^{00} &= \sum_{n=1}^{\infty} (-1)^{n+1}\frac{(n+1)^{2}M^{n}}{r^{n}}\,,
  \\
  \tilde{\mathsf{h}}^{ij} &= \sum_{n=1}^{\infty} \frac{M^{2n}x^{i}x^{j}}{r^{2n+2}}\,.
\end{aligned}\label{}
\end{equation}
As a first step, we will show by induction that the corresponding off-shell currents are given by
\begin{equation}
\begin{aligned}
  \tilde{J}^{00}_{\ord{n}|\boldsymbol{\ell}} &= 
  (-1)^{n+1}\frac{(n+1)^{2}M^{n}\pi^{\frac{d}{2}} \Gamma[\frac{d-n}{2}]}{2^{n-d} \Gamma [\frac{n}{2}] } \frac{1}{|\boldsymbol{\ell}|^{d-n}}\,,
  \\
  \tilde{J}^{ij}_{\ord{2n}|\boldsymbol{\ell}} &= 
  \frac{M^{2n} \pi^{\frac{d}{2}} 2^{d-2n}\Gamma[\frac{d-2n}{2}]}{2\Gamma[n+1] |\boldsymbol{\ell}|^{d-2n}} \Bigg[
  	\delta^{ij}{-} \frac{(d{+}2n)\ell^{i}\ell^{j}}{|\boldsymbol{\ell}|^{2}} \Bigg]\,,
  \\
  J^{00}_{\ord{n}|\boldsymbol{\ell}} &= 
  \frac{8M^{n}\pi^{\frac{d}{2}} \Gamma \left[\frac{d-n}{2}\right]}{2^{n-d} \Gamma [\frac{n}{2}] } \frac{1}{|\boldsymbol{\ell}|^{d-n}}\,, 
  \qquad
  \text{for~} n\geq3
\end{aligned}\label{currents_expression}
\end{equation}
%
%
%
%
%
%
We begin by expressing the 00-component of the Einstein tensor in terms of the new variables.
%
From this we find that $\mathsf{h}^{00}_{\ord{n}}$ ($n\geq4$) satisfies
\begin{eqnarray}
  \Delta \mathsf{h}^{00}_{\ord{n}}\!\!\! &=&\!\!\! \partial_{i} \Bigg[
    \mathsf{h}^{ij}_{\ord{2}} \partial_{j} \mathsf{h}^{00}_{\ord{n-2}}
  + \tilde{\mathsf{h}}{}^{kl}_{\ord{n-2}} \partial^{i} \mathsf{h}^{kl}_{\ord{2}} 
  - \tilde{\mathsf{h}}{}^{kl}_{\ord{n-4}} \partial_{j} \mathsf{h}^{kl}_{\ord{2}} \mathsf{h}^{ij}_{\ord{2}}
  \cr
  &&\quad 
  - \!\sum_{m=1}^{n-1} \tilde{\mathsf{h}}^{00}_{\ord{n-m}} \Big(
  	   \partial^{i} \mathsf{h}^{00}_{\ord{m}} 
  	 - \partial_{j} \mathsf{h}^{00}_{\ord{m-2}}\mathsf{h}^{ij}_{\ord{2}}
  	 \Big)
  \Bigg]\,.
\label{}\end{eqnarray}
We recast this in terms of the off-shell currents whose 00-components we write as $ J^{00}_{\ord{n}|-\boldsymbol{\ell}} = \mathcal{E}^{[1]}_{\ord{n}|-\boldsymbol{\ell}} - \mathcal{E}^{[2]}_{\ord{n}|-\boldsymbol{\ell}}$,
%
%
with
\begin{equation}
\begin{aligned}
  \mathcal{E}^{[1]}_{\ord{n}|-\boldsymbol{\ell}_{1}} &= 
  \frac{-\ell^{i}_{1}}{|\boldsymbol{\ell}_{1}|^{2}} 
  \Big(
  	  X^{i}_{\ord{n}|-\boldsymbol{\ell}_{1}} 
  	- Y^{i}_{\ord{n}|-\boldsymbol{\ell}_{1}}
  \Big)\,,
  \\
  \mathcal{E}^{[2]}_{\ord{n}|-\boldsymbol{\ell}_{1}} &= 
  \frac{-\ell^{i}_{1}}{|\boldsymbol{\ell}_{1}|^{2}} \int_{\ell_{2}}
  \Big(
  	  X^{j}_{\ord{n-2}|-\boldsymbol{\ell}_{12}}
  	- Y^{j}_{\ord{n-2}|-\boldsymbol{\ell}_{12}}
  \\&\qquad\qquad\qquad
  	+ \ell^{j}_{12} J^{00}_{\ord{n-2}|-\boldsymbol{\ell}_{12}}
  \Big) J^{ij}_{\ord{2}|\boldsymbol{\ell}_{2}}\,,
\end{aligned}\label{E12even}
\end{equation}
where
\begin{equation}
\begin{aligned}
  X^{i}_{\ord{n}|-\boldsymbol{\ell}_{1}} &= \int_{\boldsymbol{\ell}_{2}}\ell^{i}_{2} \sum_{m=1}^{n-1} \tilde{J}^{00}_{\ord{n-m}|-\boldsymbol{\ell}_{12}} J^{00}_{\ord{m}|\boldsymbol{\ell}_{2}}\,,
  \\ 
  Y^{i}_{\ord{n}|-\boldsymbol{\ell}_{1}} &= \int_{\boldsymbol{\ell}_{2}}\ell^{i}_{2} \tilde{J}^{kl}_{\ord{n-2}|-\boldsymbol{\ell}_{12}} J^{kl}_{\ord{2}|\boldsymbol{\ell_{2}}}\,.
\end{aligned}\label{}
\end{equation}
%

We start the proof by induction at $n=3$ where we have explicitly verified that 
\eqref{currents_expression} holds. Assuming the currents are given by \eqref{currents_expression} 
up to the $(n-1)$-th order, we next show that it also holds for the $n$-th order term.
Dividing into odd and even $n$, and assuming that \eqref{currents_expression} holds up to order
$n-1$, we find
\begin{equation}
\begin{aligned}
  \mathcal{E}^{[1]}_{\ord{2n}|-\boldsymbol{{\ell}}} &= 
  \frac{M^{2n}\pi^{\frac{d}{2}} 2^{d-2 n+3}  \Gamma[\frac{d}{2}-n]}{\Gamma[n]} \frac{1}{|\boldsymbol{\ell}|^{d-2 n}}\,,
  \\
  \mathcal{E}^{[2]}_{\ord{2n}|-\boldsymbol{{\ell}}} &= 0\,,
\end{aligned}\label{}
\end{equation}
and
\begin{equation}
\begin{aligned}
  \mathcal{E}^{[1]}_{\ord{2n+1}|-\boldsymbol{{\ell}}} &= 
  \frac{M^{2n+1}\pi ^{\frac{d}{2}} (4 n+1) 2^{d -2 n} \Gamma [\frac{d -2 n-1}{2}]}
{\Gamma [n+\frac{3}{2}]|\boldsymbol{\ell}|^{d -2 n-1}}\,,
  \\
  \mathcal{E}^{[2]}_{\ord{2n+1}|-\boldsymbol{{\ell}}} &= 
  \frac{M^{2n+1}\pi ^{\frac{d}{2}} 2^{d -2 n} \Gamma [\frac{d -2 n-1}{2}]}
{\Gamma[n+\frac{3}{2}]|\boldsymbol{\ell}|^{d -2 n-1}}\,.
\end{aligned}\label{}
\end{equation}
For both odd and even $n$ these terms combine to yield eq. \eqref{currents_expression}. Going to four 
space-time dimensions, we find
\begin{equation}
\mathsf{h}^{00}_{\ord{n}} = \frac{8M^{n}}{r^{n}}\,,
\end{equation}
for all $n \geq 3$.
The hardest parts to compute are the $ij$-components. We find it convenient to first introduce
\begin{equation}
\begin{aligned}
   Z^{ij}{}_{kl} &= \mathfrak{g}^{im}\partial_{m}\mathfrak{g}^{jn} \partial_{k}\tilde{\mathfrak{g}}_{nl} \,,
 \\
 W^{i}{}_{l} &= \mathfrak{g}^{ki} \partial_{k}\mathfrak{g}^{00} \partial_{l} \tilde{\mathfrak{g}}_{00} \,,
 \\
  d^{i} &= \frac{1}{2(d-1)} \mathfrak{g}^{ij}\mathfrak{g}^{\rho \sigma} \partial_{j} \tilde{\mathfrak{g}}_{\rho \sigma}\,.
\end{aligned}\label{}
\end{equation}
%
\begin{equation}
\begin{aligned} 
  \Delta \mathsf{h}^{ij}\! &= 
    \mathsf{h}^{kl} \partial_{k}\partial_{l}\mathsf{h}^{ij} 
  {-} \partial_{l} \mathsf{h}^{ki} \partial_{k} \mathsf{h}^{l j}
  {+} 2(d{-}1) d^{i} d^j
  {+} 2 \mathfrak{g}^{ij} \partial_{k}d^{k}
  \\ & \quad
  {+} Z^{k(i}{}_{k}{}^{j)}
  {-} 2 Z^{(i|k|}{}_{k}{}^{j)} 
  {+} \frac{1}{2} Z^{(i|k|j)}{}_{k}
  {+} \frac{1}{2} W^{ij}
  \\&\quad
  - \frac{1}{2} \Big[
  	  2 Z^{k(i}{}_{kl} 
  	{-} 4 Z^{(i|k|}{}_{kl} 
  	{+} Z^{(i|k|}{}_{lk}
  	{+} W^{i}{}_{l}
  \Big] \mathsf{h}^{j)l}\,.
\end{aligned}
\label{h_ij_eom}\end{equation}
%
%
%
%
%
%
%
In terms of currents, we have
\begin{equation}
    d^{i}_{\ord{n}|-\boldsymbol{\ell}_{1}} {=} 
    \frac{i}{2(d-2)} \bigg[
      \mathcal{D}^{i}_{\ord{n}|-\boldsymbol{\ell}_{1}} 
  	{-} \int_{\boldsymbol{\ell}_{2}} \mathcal{D}^{j}_{\ord{n-m}|-\boldsymbol{\ell}_{12}} J^{ij}_{\ord{m}|\boldsymbol{\ell}_{2}}
    \bigg]\,,
\label{}\end{equation}
where
\begin{equation}
  \mathcal{D}^{i}_{\ord{n}|-\boldsymbol{\ell}_{1}} 
  {=} \bigg[
    \ell^{i}_{1} J^{00}_{\ord{n}|-\boldsymbol{\ell}_{1}}
  {-} \ell^{i}_{1} J^{kk}_{\ord{n}|-\boldsymbol{\ell}_{1}}
  {+} X^{i}_{\ord{n}|-\boldsymbol{\ell}_{1}}
  {+} Y^{i}_{\ord{n}|-\boldsymbol{\ell}_{1}}
  \bigg]\,.
\label{}\end{equation}
%
Iterating, we find
\begin{equation}
\begin{aligned}
  \mathcal{D}^{i}_{\ord{2n}|-\boldsymbol{\ell}_{1}} &= 
  \frac{\pi ^{\frac{d}{2}} 2^{d-2 n+1} \Gamma[\frac{d}{2}-n]}{\Gamma[n+1]}	 \frac{\ell^{i}}{|\boldsymbol{\ell}|^{d-2 n }}\,,
  \\
  \mathcal{D}^{i}_{\ord{2n+1}|-\boldsymbol{\ell}_{1}} &= 
  -\frac{\pi^{\frac{d}{2}} 2^{d -2 n} \Gamma[\frac{d-1}{2}-n]}{\Gamma[n+\frac{3}{2}]} \frac{\ell^{i}}{|\boldsymbol{\ell}|^{d-2 n-1}}\,,
\end{aligned}\label{}
\end{equation}
%
%
%
so that 
\begin{equation}
\begin{aligned}
  d^{i}_{\ord{1}|-\boldsymbol{\ell}_{1}} &= 2^{d -1} \pi^{\frac{d -1}{2}} |\boldsymbol{\ell}_{1}|^{1-d } \Gamma[\tfrac{d -1}{2}] \ell^{i}\,,
  \\
  d^{i}_{\ord{2}|-\boldsymbol{\ell}_{1}} &= -2^{d-3} \pi ^{\frac{d}{2}} |\boldsymbol{\ell}_{1}|^{2-d} \Gamma[\tfrac{d}{2}-1] \ell^{i}\,,
  \\
  d^{i}_{\ord{n}|-\boldsymbol{\ell}_{1}} &= 0\,,  ~~~ \text{for}~n \geq3\,.
 \end{aligned}\label{}
\end{equation}
Similarly, we find
\begin{equation}
W^{ij}_{\ord{3}|-\boldsymbol{\ell}_{1}} = \frac{\Gamma[\tfrac{d -5}{2}] \big(|\boldsymbol{\ell}_{1}|^2 \delta^{ij}-(d -5) \ell_{1}^{i} \ell_{1}^{j}\big)}{2^{-d-1}15 \pi^{\frac{1-d}{2}}}  |\boldsymbol{\ell}_{1}|^{3-d }\,,
\end{equation}
and, for $n > 2$,
\begin{equation}
\begin{aligned}
  W^{ij}_{\ord{2n}|-\boldsymbol{\ell}} &= 
  \frac{5 \pi^{\frac{d}{2}} \Gamma[\frac{d-2n-2}{2}]}{2^{2 n+1-d}\Gamma [n{+}2]} 
  \frac{(2n{+}2{-}d) \ell^{i} \ell^{j}{-}|\boldsymbol{\ell}|^2 \delta^{ij}}{|\boldsymbol{\ell}|^{d-2 n}}\,,
  \\
  W^{ij}_{\ord{2n+1}|-\boldsymbol{\ell}} &= 
  \frac{\pi^{\frac{d}{2}} \Gamma[\frac{d {-}2 n{-}3}{2}]}{2^{2n-d} \Gamma[n{+}\frac{5}{2}]} \frac{(d {-}2 n{-}3) \ell^{i} \ell^{j}{-}|\boldsymbol{\ell}|^2 \delta^{ij}}{|\boldsymbol{\ell}|^{d -2n-1}}\,.
\end{aligned}\label{}
\end{equation}
%
%
%
Substituting these results all terms with auxiliary fields $d^i$ and $W^{ij}$ cancel for $n \geq 4$. Finally, also all terms with the $Z$-fields combine to cancel separately. Thus, in total,
\begin{equation}
\mathsf{h}^{ij}_{(n)} = 0\,, ~~~ n \geq 3 ~. 
\end{equation}

To summarize this part, we have shown to all orders in perturbation theory that $h^{\mu\nu}$ 
satisfies the expansions \eqref{metric_components} and \eqref{Laurent_series}. The expansion for
$h^{00}$ is convergent for $GM/r < 1$ and sums to the correct expression in that region while the
expansion for $h^{ij}$ truncates at order $G^2$, in agreement with eq. \eqref{metric_components}.

As is well known, the harmonic gauge condition does not determine uniquely the standard harmonic-gauge metric \eqref{metric_components} due to a residual choice of coordinate freedom within the harmonic gauge itself \cite{Fromholz:2013hka}. This prompts the question as to how such a residual freedom should manifest itself within our recursive solution that seems to uniquely provide the standard form \eqref{metric_components}. The issue is resolved by noticing that the perturbative solution requires regularization of the pertinent loop integrals beyond the first few leading orders. Indeed, up to order $G^{2}$ there is no need for regularizing the integrals and there is correspondingly no additional coordinate freedom of the harmonic-gauge metric \cite{Fromholz:2013hka}. At order $G^3$ we need to regularize the loop integrals. Simultaneously, we encounter constant numerators in the loop integrands. Those terms lead to (unregularized) integrals that are scale-free and they are set to zero in dimensional regularization. Other equally valid regularization schemes may set those integrals to other constant values, and thus introduce new scheme-dependent constants in the metric. By iteration, such constants will enter at all higher orders as well. This is in agreement with the arbitrariness stemming from residual coordinate freedom in this gauge.

\section{Conclusion}

Applying the perturbiner method to Einstein gravity we have shown how
to derive the harmonic-gauge metric of a Schwarzschild black hole to all orders in perturbation theory. 
This is the first example of a post-Minkowskian expansion in general relativity that has been
carried through to all orders and which therefore teaches us about the convergence of such
expansions. We find that the series is convergent within the expected range based on the closed expression of eq. \eqref{metric_components}. What makes this all-order result possible in spite of the 
overwhelming complexity of perturbative gravity is the iterative systematics of the
perturbiner expansion for gravitational equations of motion, combined with recursive structures
of the involved loop integrals. In addition, complications arising from the
perturbative expansion of the Einstein-Hilbert action can be controlled by a doubling
of degrees of freedom at intermediate steps. This highlights the fact that also the weak-field
expansion of the action contains iterative structures, here inherited from the geometric series relation between metric and inverse metric, an observation made earlier in ref. \cite{Cheung:2017kzx}. The perturbiner approach used here is an efficient
recursive technique for solving classical equations of motion. The resulting expressions resemble closely the classical parts of the loop expansion of perturbative gravity, and we have a few times used the language of the loop expansion even though all
integrations here arise only from the fact that the Green function technique works well through its Fourier transform.
The way classical contributions can arise in General Relativity from the loop expansion was clearly explained in
ref. \cite{Holstein:2004dn}.
In a broader perspective, one may hope that the perturbiner approach to post-Minkowskian gravity \cite{Cho:2021nim,Lee:2022aiu,Cho:2022faq,Adamo:2023cfp,Lee:2023zuu,Cho:2023kux} could lead to related simplifications for the general relativistic two-body problem.

\smallskip

\begin{acknowledgments}
{\sc Acknowledgments:}~ We thank Thibault Damour for discussions regarding convergence properties of the post-Minkowskian expansion. The work of P.H.D. is supported in part by DFF Grant No. 0135-00089A, and the work of K.L. has been supported in part by appointment to the JRG Program at the APCTP through the Science and Technology Promotion Fund and Lottery Fund of the Korean Government. KL is also supported by the National Research Foundation of Korea(NRF) grant funded by the Korean government(MSIT) RS-2023-00249451 and the Korean Local Governments of Gyeongsangbuk-do Province and Pohang City.

\end{acknowledgments}

\end{document}